# Electron transport and thermoelectric properties of layered perovskite $LaBaCo_2O_{5.5}$

Asish K. Kundu,[1] B. Raveau, V. Caignaert, E.-L. Rautama and V. Pralong

CRISMAT Laboratory, ENSICAEN UMR6508, 6 Boulevard Maréchal Juin, Cedex 4, Caen-14050, France

## Abstract

We have investigated the systematic transport properties of the layered 112-type cobaltite $LaBaCo_2O_{5.5}$ by means of electrical resistivity, magnetoresistance, electroresistance and thermoelectric measurements in various conditions. In order to understand the complex conduction mechanism of $LaBaCo_2O_{5.5}$, the transport data have been analyzed using different theoretical models. The system shows semiconductor-semiconductor like transition ($T_{SC}$) around 326K, corresponding to ferromagnetic transition and in the low temperature region resistivity data follows the Mott's variable range hopping model. Interestingly, near and below the room temperature this compound depicts significant change in electro- and magnetoresistance behavior, the latter one is noteworthy near the magnetic phase boundary. The temperature dependence of thermopower, S(T), exhibits p-type polaronic conductivity in the temperature range of $60K \leq T \leq 320K$ and reaches a maximum value of ~303μV/K (at 120K). In the low temperature AFM region, the unusual S(T) behavior, generally observed for the cobaltite series $LnBaCo_2O_{5.5}$ (Ln = Rare Earth), is explained by the electron magnon scattering mechanism as previously described for perovskite manganites.

(All figures in this article are in colour only in electronic version)

[1]Corresponding author *E-mail*: asish.kundu@ensicaen.fr/asish.k@gmail.com.

## 1. INTRODUCTION

The layered perovskite cobaltites of 112-type $LnBaCo_2O_{5.5}$ (Ln = Rare Earth) have been of great interest due to their rich physical properties and interesting structural phenomena associated with them [1-5]. In the layered cobaltites where the oxygen stoichiometry is "5.5", the average valency of cobalt ion is $Co^{3+}$. This is particularly interesting because of the ordering of $Co^{3+}$ ions in two different crystallographic sites corresponding to pyramidal and octahedral oxygen coordination [1,3,5]. Although, the series has been investigated in details for almost all lanthanide elements of the periodic table, interestingly there is no such report [1,2] on the first member of this series, i.e. for the lanthanum phase $LaBaCo_2O_{5.5}$. Importantly, the $LnBaCo_2O_{5.5}$ systems possess a layered crystal structure which consists of layers $[LnO_x]$-$[CoO_2]$-$[BaO]$-$[CoO_2]$ alternating along the c-axis [1,3,5]. The ordering of $Ln^{3+}$ and $Ba^{2+}$ ions is favourable if the size difference is large between the cations, hence smaller size $Ln^{3+}$ easily form a layered structure. In contrast, for $La^{3+}$ ions the size difference with $Ba^{2+}$ is smaller, and as a result the disordered cubic perovskite [2] is more stable and it requires special conditions to synthesize the 112-type layered structure. Very recently we have studied $LaBaCo_2O_{5.5}$ in terms of neutron diffraction, electron microscopy and magnetic studies, which shows that the structure is 112-layered orthorhombic with G-type antiferromagnetic (AFM) ordering [6]. Therefore, in this article we have investigated in details the transport properties of $LaBaCo_2O_{5.5}$. Interestingly, the compound shows an electronic transition ($T_{SC}$) between two semiconducting states above room temperature $T_{SC}$ ~326K, which coincides with paramagnetic (PM) to ferromagnetic (FM) transition. We have also carried out thermopower (Seebeck coefficient), S(T), and thermal conductivity, κ(T), measurements to investigate the electron transport properties of the system below $T_{SC}$.

The latter are sensitive to the magnetic and electrical nature of charge carriers (hole/electron) and can give some valuable information, which are absent in the magnetotransport measurements. Additionally, the S(T) data is less affected due to the presence of grain boundaries, which often complicates the ρ(T) data interpretation for polycrystalline samples. We noticed that, at room temperature the system shows a relatively large positive value of the thermoelectric power (91 μV/K), while it has a low resistivity (~1.14 Ω cm at 300K). Therefore, we have calculated power factor [7], $S^2\sigma$, as well as the figure of merit (ZT) as characterized for thermoelectric materials by $ZT = S^2\sigma T/\kappa$, where S, σ(~1/ρ), κ and T represent thermopower, electrical conductivity, thermal conductivity and absolute temperature, respectively.

## 2. EXPERIMENTAL PROCEDURE

The polycrystalline sample was synthesized by means of a soft-chemistry method [6]. Phase purity, chemical analysis and oxygen stoichiometry were carried out for the sample as discussed in Ref. 6. The magnetization, resistivity and thermopower measurements were carried out with a Quantum Design physical properties measurements system (PPMS). The electrical measurements were carried out on a rectangular-shaped ($7.10 \times 2.85 \times 2.14 mm^3$) sample by a standard four-probe method and a home made sample holder was used for thermoelectric power measurements (using a steady state technique).

## 3. RESULTS AND DISCUSSION

Figure 1 shows the temperature dependence of electrical resistivity, ρ(T), and magnetization, M(T), for $LaBaCo_2O_{5.5}$ in the temperature range of 10-400K. The ρ(T) data is collected during heating and cooling cycle of the measurements in the presence and absence of the external magnetic field (7 Tesla(T)). The zero field ρ(T) curve shows a significant change in slope corresponding to the semiconductor-semiconductor transition ($T_{SC}$) around 326 K (figure 1(a)). This type of transition is previously reported for the other series of $LnBaCo_2O_{5.5}$ [1-5], and referred to as insulator-metal transition ($T_{IM}$) albeit the true nature of this transition is semiconducting to semiconducting type. For the present $LaBaCo_2O_{5.5}$ system, in contrast to a metallic behavior, the slope of the resistivity curve (dρ/dT) is negative above the transition temperature ($T > T_{SC}$), similar to the $YBaCo_2O_{5.45}$ system reported by Pautrat et al [4]. We have also studied (data not shown) carefully, the ρ(T) behavior for $LnBaCo_2O_{5.5}$ (where Ln = Nd, Sm, Eu, Gd, Dy and Y), near and above the transition temperature and noticed that the slope (dρ/dT) is negative for all of them. Furthermore, for $LaBaCo_2O_{5.5}$ in an applied field of 7 T we did not observe any significant change in the resistivity behavior through out the temperature range, except a slight decrease in the magnitude below $T_{SC}$. It is observed (from figure 1) that, the electronic and magnetic transition temperatures for $LaBaCo_2O_{5.5}$ are almost the same ($T_C$ and $T_{SC} \approx 326$K), in contrast to other perovskite cobaltites $LnBaCo_2O_{5.5}$, which exhibit a large shift between $T_C$ and $T_{IM}$ [1,2]. We have re-investigated the magnetic properties of $LaBaCo_2O_{5.5}$, the zero field cooled (ZFC) and field cooled (FC) magnetization data (H = 0.5 T) is shown in figure 1(b), which depicts two characteristic transitions. Moreover, the spin state transition is seems to be absent in the whole temperature regime (most probably $Co^{3+}$ is in the intermediate spin state throughout [6]). Hence, the "spin-state" problem is not of prime important for our present study. The general feature of the present study is the significant thermomagnetic irreversibility under higher field, below the FM transition $T_C$ and the presence of weak magnetic transition at low temperature. Hence, the sample is magnetic-semiconductor below $T_{SC}$ and the resistivity increases exponentially with decreasing temperature. The high temperature ρ(T) behavior is weakly temperature dependent and is not affected by the magnetic field or measurements cycling process. As can be seen from the inset figure 1(a), the ρ(T) curves for cooling (triangle symbol) and heating (circle symbol) cycles, show only little divergence below $T_{SC}$ and merge above this temperature.

The resistivity jump, between two semiconducting states coincides with the PM to FM-like transition described in the previous section. The semiconductor or insulator like transport in perovskite cobaltites can be characterized by three possible models [5,8] namely thermal activation (TA): $\log \rho \propto 1/T$, Efros-Shklovskii type hopping (ESH): $\log \rho \propto T^{-1/2}$ and Mott's variable range hopping (VRH): $\log \rho \propto T^{-1/4}$. To understand the transport mechanism for $LaBaCo_2O_{5.5}$, the data have been analyzed based on these models. The zero field resistivity data are plotted in figure 2, which show that the VRH model fits better than the other two models in the temperature range of $40K \leq T \leq 270K$, and is consistent with the previous studies on perovskite cobaltites [5,8]. This type of transport mechanism is typical for disordered systems, where the charge carriers move by hopping between localized states. Taskin et al [5] have described the formation of localized states in terms of oxygen defects in $GdBaCo_2O_{5.5}$, which inevitably generate electrons or holes in the $CoO_2$ planes. Moreover, we apply ESH and TA models, which are expected to describe the dominant conduction process. The transport mechanism is very complicated in the whole temperature region and does not satisfy completely any of the above models. It can be seen from the inset of figure 2, that near or above the room temperature they diverge from experimental data point. From the semiconducting region ($T \leq 250K$) of the TA model we have calculated the approximate activation energy ($E_a$) of around 21 meV, which is a typical value for a narrow band gap insulator, yet less than the value reported for $GdBaCo_2O_{5.5}$ [5].

Figure 3, shows the magnetic field dependent isothermal magnetoresistance (MR) effect for $LaBaCo_2O_{5.5}$ at five different temperatures. Unfortunately, in the case of temperature dependent (with magnetic field) resistivity data, we did not observe any major change in the ρ(T) curve although the magnetization behavior reveals some kind of FM and AFM ordering. The charge transport for this kind of system is expected to be very sensitive, due to the presence of FM and AFM state and the external magnetic fields readily induces an MR effect by affecting the subtle balance between FM-AFM phases. Therefore, we have studied the isothermal MR effect at different temperatures and a clear magnetic field dependent change in the MR behavior is noticed below $T_{SC}$. The MR value is calculated [9] as, MR (%) = [{ρ(7)-ρ(0)}/ρ(0)]x100, where ρ(0) is the sample resistivity at 0 T and ρ(7) in an applied field of 7 T. The highest negative MR value is obtained at 245 K, around -5 % in an applied field of 7 T, and near $T_N$ the value is only -1.6 % (at 300K). At low temperatures, the MR values are only -2.8 % (at 150K) and -2.5 % (at 50K) respectively. The evidence of negative MR at low temperatures, similar to the tunnelling MR observed usually in polycrystalline samples for hole doped cobaltites, is considered to be related to spin-dependent scattering at grain boundaries. Nevertheless, in the present case the highest MR value is noticed near the FM-AFM phase boundary, so that the grain boundary effect can be ignored and considered to be as an intrinsic effect of the sample. Interestingly, the magnetic field dependent isothermal MR behavior at 245 K exhibits an irreversible effect similar to those of isothermal magnetization, M(H), behavior (see inset figure 3), which is also present in 300 K isothermal MR and M(H) data. The peak in the isothermal MR data occurs

around the coercive field value, which corresponds to the state of maximum disorder in the orientation of the neighbouring magnetic spins. Hence, the field dependent MR data, which is indirectly related to the alignment between magnetic spins, reaches a maximum value. This effect is prominent for 300K data, compared to 245K as evidenced from the inset figure 3 (dotted vertical lines). Additionally, the isothermal MR data exhibit hysteresis effects below room temperature, which resembles the “butterfly-like” feature, although the effect is rather weak at low temperature (50K). It is clear from the obtained data that the MR effect is anisotropic near the FM-AFM phase transition (studied at 245, 275 and 300 K). The strong anisotropic “butterfly-like” feature appears only near the phase boundary. Hence, the occurrences of anisotropic MR behavior nearly at similar temperatures as those for magnetic field variation isothermal M(H) studies suggest the strongly correlated nature of field-induced magnetic and electronic transitions, as reported earlier for nanoscale ordered $LaBaCo_2O_6$ [9].

Furthermore, we have carried out the isothermal electric-field effect at different temperatures akin to MR effect with a view to understand the transport mechanism below $T_{SC}$. Figure 4(a) shows the current variation of normalized resistance, $R(I) / R(I_1)$; where $R(I)$ is the electrical resistance at 100 mA and $R(I_1)$ for 1mA current, at four different temperatures. At 300K, the resistance ratio initially decrease rapidly with increasing current (≈10 mA), but eventually for higher current the resistance decreases gradually. A similar feature is also observed at low temperatures (studied at 250 and 200K). The latter may be due to the low thermal conductivity and higher resistivity of $LaBaCo_2O_{5.5}$ sample at low temperature ($\kappa \sim 2$ W $K^{-1}m^{-1}$ and $\rho \sim 14\ \Omega$ cm at 200K), which may suggest another argument like Joule heating effect. This effect should indeed lead to a gradual decrease of the electrical resistance with increasing current due to the enhancement of dissipation power (inside the bulk sample), due to the poor thermal conductivity as pointed out by Mercone et al [10]. However, the observed electric field effects in the present case may be intrinsic near the room temperature since the thermal conductivity is relatively higher. Therefore, we have carefully investigated the current (I)-voltage (V) characteristics at same temperatures as shown in figure 4(b). It is observed that the linear Ohmic behavior is obtained for I-V curves at 390K and 300K (for low current, I ≤ 35 mA), respectively. In contrast, for 250K (200K) data, the behavior is nonlinear beyond the current value of ~20 (15) mA, which implies that in the low temperature semiconducting regime the dominant effect may be due to Joule heating [10]. Hence, from the obtained data (figure 4) it is clear that the electroresistance effect, at 300K for low current (I ≤ 35 mA i.e. in the linear region of I-V curve), is due to the intrinsic nature of the $LaBaCo_2O_{5.5}$ sample, whereas at low temperatures the dominant effect is indeed Joule heating. Further investigations are required to confirm this interesting physical phenomenon.

In order to get more insight into the nature of the conduction mechanism below the $T_{SC}$, we have studied the thermoelectric power measurements. This is a simple and sensitive technique to investigate the scattering mechanism in electronic conduction, as discussed earlier. The temperature dependence of thermopower, S(T), for cooling and heating cycles is shown in figure 5(a). It is observed that, the S(T) value is positive below $T_{SC}$ and with decreasing temperature the value increases rapidly and reaches a maximum value of ~303 μV/K at around 120K (referred as $T_P$). Decreasing the temperature further beyond $T_P$, the S(T) value decreases continuously and due to the limitation of our instrument we were unable to collect data below 60K. The temperature denoted by an arrow in the figure represents the broad peak value $T_P$

~120K for the present system. We have also investigated the S(T) evolution for $EuBaCo_2O_{5.5}$ (data not shown), which depicts similar features with $T_P$ ~ 90 K. From the literature values we have plotted $T_P$ as a function of different rare earth's size (i.e. for $NdBaCo_2O_{5.5}$, $GdBaCo_2O_{5.5}$ and $HoBaCo_2O_{5.5}$ the approximate $T_P$ [obtained from S(T) behavior] values are 105, 88 and 70K respectively) [3,5]. We obtain a linear decrease of $T_P$ with the rare earth size (data not shown). It is well known that, the average cation size plays a crucial role in electronic conduction due to the change in electronic band width. In the present systems, the decrease in $T_P$ with the rare earth size is related to the decrease in band width or increasing the energy band gap.

The S(T) behavior implies that the $LnBaCo_2O_{5.5}$ systems show a semiconducting type behavior of the thermoelectric power down to $T_P$ akin to resistivity behavior (figure 1(a)) and below this temperature the S(T) behavior is complex, decreasing abruptly at low temperature. This type of behavior is quite unusual for semiconducting thermoelectric materials. In fact, with decreasing temperature, the S(T) value should increase, due to trapping or localization of charge carriers. Unlike the resistivity the S(T) behavior is unusual (60K $\leq$ T $\leq$ 320K) and there is no general explanation till date. Following the general approach in analyzing the semiconducting behavior we have plotted the S(T) data in the $T^{-1/n}$ scale similar to $\rho(T)$. Since for semiconductors the S(T) is expected to be linear in $T^{-1}$ behavior (according to TA model) or to follow either of the described hopping models similar to resistivity behavior [5,8]. The plots show that (not shown) the fitting is very poor even for short temperature range. Therefore, the S(T) data can not be described by the previous mentioned models and one observes p-type conductivity throughout the temperature range (60-320K). Many authors have already explained the S(T) behavior at higher temperature and their sign reversal near the $T_{SC}$ or $T_{IM}$ [3,5]. However, the low temperature S(T) evolution, basically the appearance of the broad maximum ($T_P$) in the metastable AFM phase and the decreasing nature with temperature (in spite of semiconducting behavior) has not been explained properly. In this respect it is important to investigate the present system below room temperature. We have analysed the obtained thermopower data using an expression $S(T) = S_0 + S_{3/2}T^{3/2} + S_4T^4$ defined by P. Mandal [11], which can be understood on the basis of electron magnon scattering (spin wave theory). In ferro- and antiferromagnets, electrons are scattered by spin wave as explained earlier for perovskite manganites [11] and hence it is expected that this theory will explain the S(T) behavior for the present system. However, the thermopower value is much higher than the manganites. The S(T) curve shows a positive curvature below $T_P$ and we notice that at low temperature the S(T) data follows a $T^{3/2}$ behavior (60-105K) and in the 120-320K range it linearly fits with the $T^4$ behavior (insets of figure 5). The corresponding fitting coefficients $S_{3/2}$ and $S_4$ are $65x10^{-3}$ $\mu V/K^{5/2}$ and $-2.7x10^{-8}$ $\mu V/K^5$ respectively and the obtained values are one order of magnitude higher than for the hole doped manganites [11]. At low temperature the second term $S_{3/2}$ will dominate over $S_4$ ($S_{3/2}>>S_4$), hence the S(T) will depict downward trend below $T_P$. Although we do not have sufficient experimental data points below the broad peak, yet the S(T) curve below $T_P$ fits linearly to the $T^{3/2}$ term (inset of figure 5(b)), as expected from the spin wave theory. Moreover, the downward feature in S(T) is present in $LnBaCo_2O_{5.5}$ cobaltites [3,5]. Hence, the broad peak at low temperature and downward trends for layered cobaltites are considered to be the result of the electron magnon scattering akin to perovskite manganite as explained by P. Mandal [11].

Temperature dependent thermal conductivity, κ(T), measurements of $LaBaCo_2O_{5.5}$ during cooling and heating cycles is shown in figure 5(b). We have plotted the data up to 250K, as beyond this temperature the κ(T) measurements are affected by thermal radiation effects. The κ(T) value increases with increasing temperature, indicating an usual phonon mediated scattering mechanism of charge carriers, which is the sum of phonon ($\kappa_p$) and electronic ($\kappa_e$) contribution [i.e. κ(T) = $\kappa_p$(T) + $\kappa_e$(T)]. Therefore, from the Wiedemann-Franz law we have calculated the electronic contribution ($\kappa_e$) at 250K, which is around ~1.5x$10^{-4}$ $WK^{-1}m^{-1}$ [whereas κ(250K) ~ 2.8 $WK^{-1}m^{-1}$]. This implies that, the lattice modulated phonon contribution ($\kappa_p$) is dominant near room temperature i. e. $\kappa_p$(T) >> $\kappa_e$(T). The temperature dependence of the power factor ($S^2\sigma$) as shown in figure 5(b), depicts a clear maximum near 270K and decreases rapidly on both sides of the temperature scale (60K ≤ T ≤ 320K). It is important to note that, this maximum is in the region of FM-AFM phase boundary. This suggests that the electron magnon scattering at low temperature strongly affects the S(T) behavior of $LaBaCo_2O_{5.5}$. The figure of merit (ZT) shows too small values close to room temperature (~$10^{-5}$) for applications (inset figure 5(b)), as expected from 112-type cobaltite systems.

## 4. CONCLUSIONS

The prime result of current investigations is the unusual S(T) behavior in the metastable AFM state and the appearance of a broad peak around 120K with a positive value of ~303 μV/K. This is explained by electron magnon scattering mechanism, which is expected to be applicable to all series of $LnBaCo_2O_{5.5}$ cobaltites at low temperature. Moreover, the magnetic field dependent isothermal MR data exhibits an irreversible feature with the highest value of MR around 5 % (at 245K and 7T) near the FM-AFM phase boundary. The peak in the MR data appears close to coercive field value (in M(H) data), suggesting maximum domain wall interactions. This implies a strongly correlated nature of the magnetic and electron transport properties. Importantly, the present sample also shows an electric field effect at room temperature, which is seems to be an intrinsic property of the system, as evidenced from the I-V characteristics.

## Acknowledgements

AKK gratefully acknowledges Dr. A. Pautrat and Dr. S. Hebert for valuable suggestions during electroresistance and thermoelectric measurements and also for carefully reading this manuscript. The authors thank the CNRS and the Ministry of Education and Research for financial support.

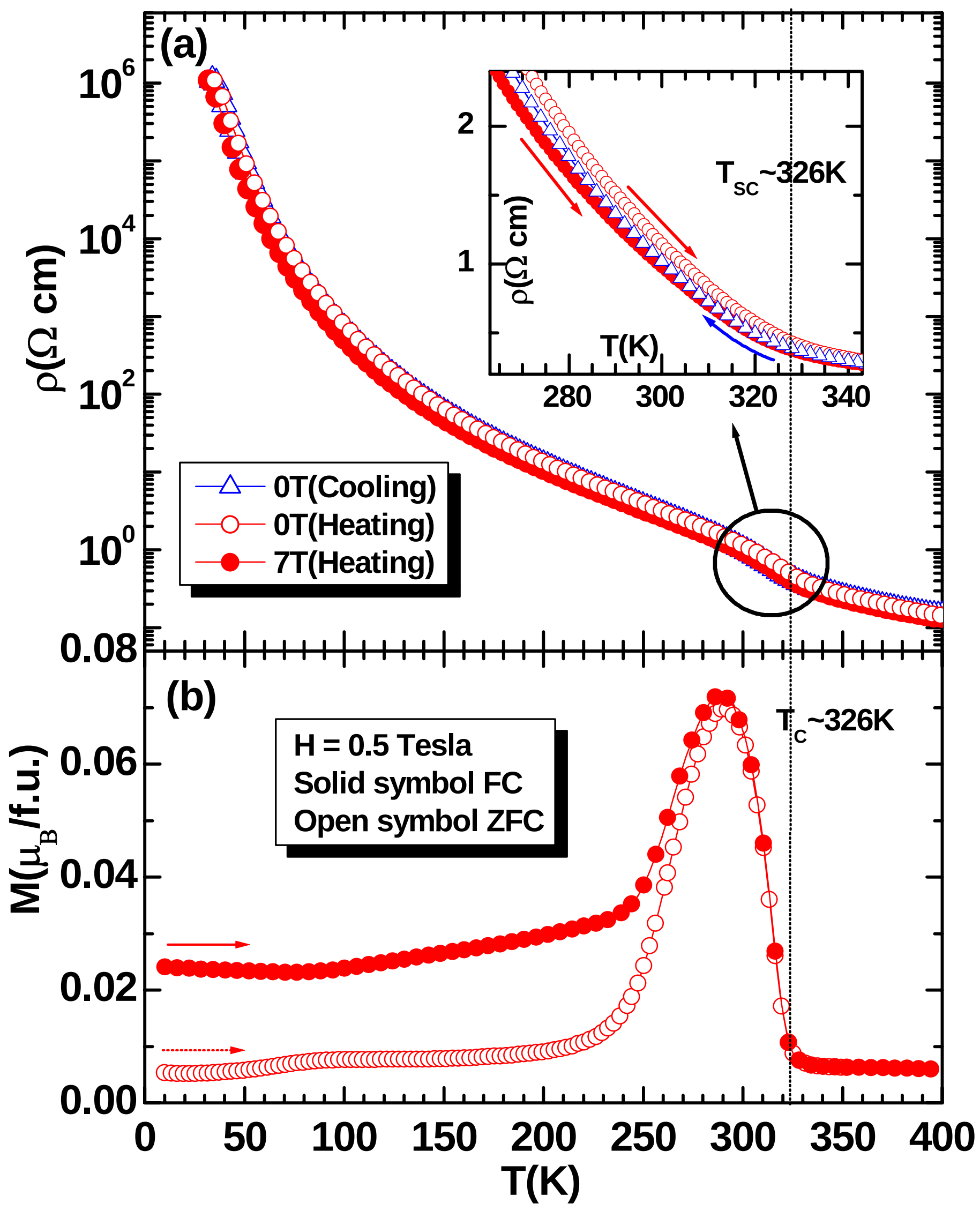

**Figure 1.** Temperature dependent physical properties for ordered $LaBaCo_2O_{5.5}$: (a) electrical resistivity, ρ(T), in the presence (solid symbol) and absence (open symbol) of magnetic field (7 Tesla) during heating (circle) and cooling (triangle) cycles (inset shows the expanded version near the transition temperature, $T_{SC}$), and (b) ZFC (open symbol), FC (solid symbol) magnetization, M(T), in an applied field of 0.5 Tesla.

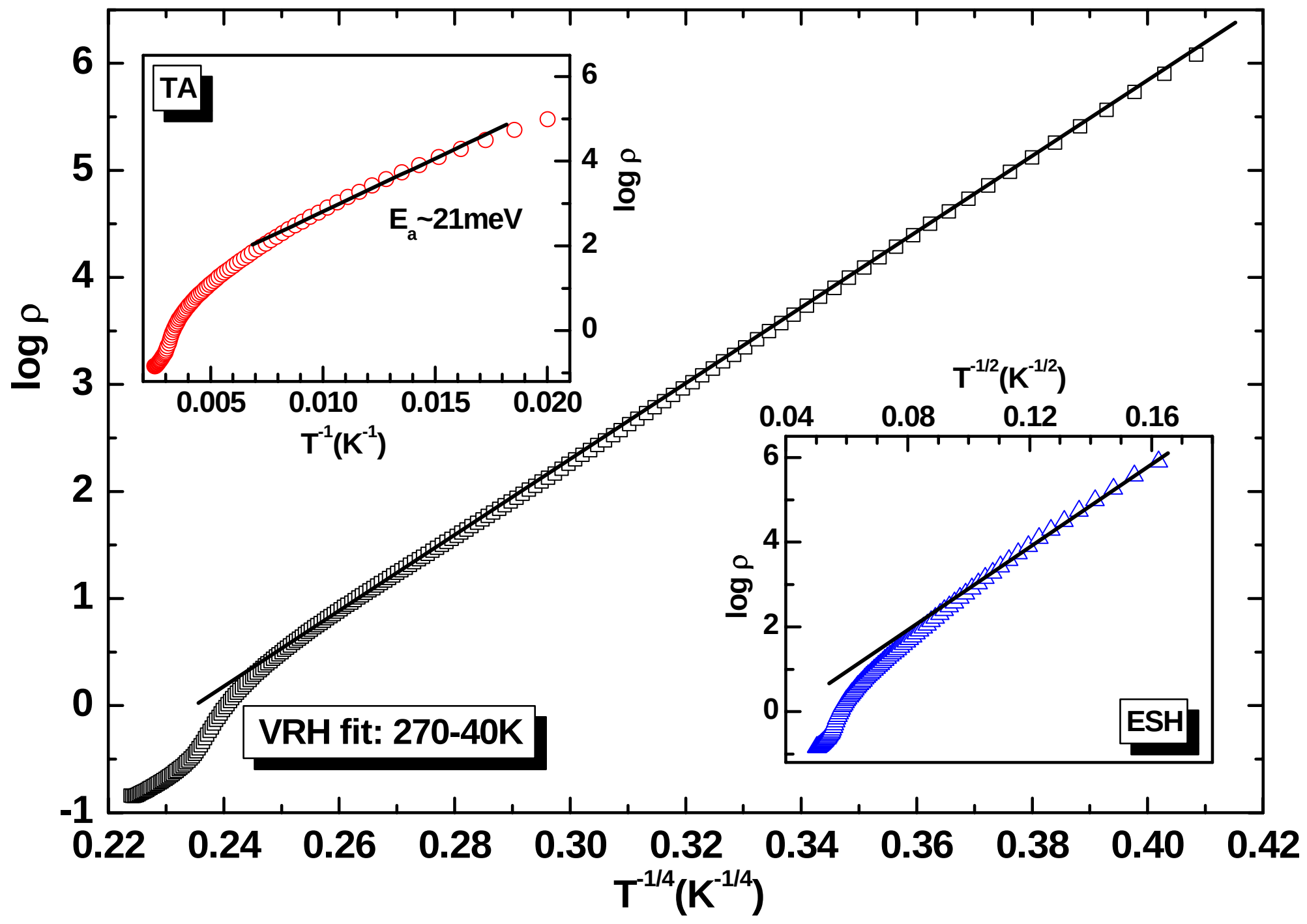

**Figure 2.** Logarithm of the resistivity versus $T^{-1/n}$ plots (where n = 1, 2 or 4) for ordered $LaBaCo_2O_{5.5}$: open symbols and solid lines represent the experimental data and apparent fit to the different hopping models as described in the text.

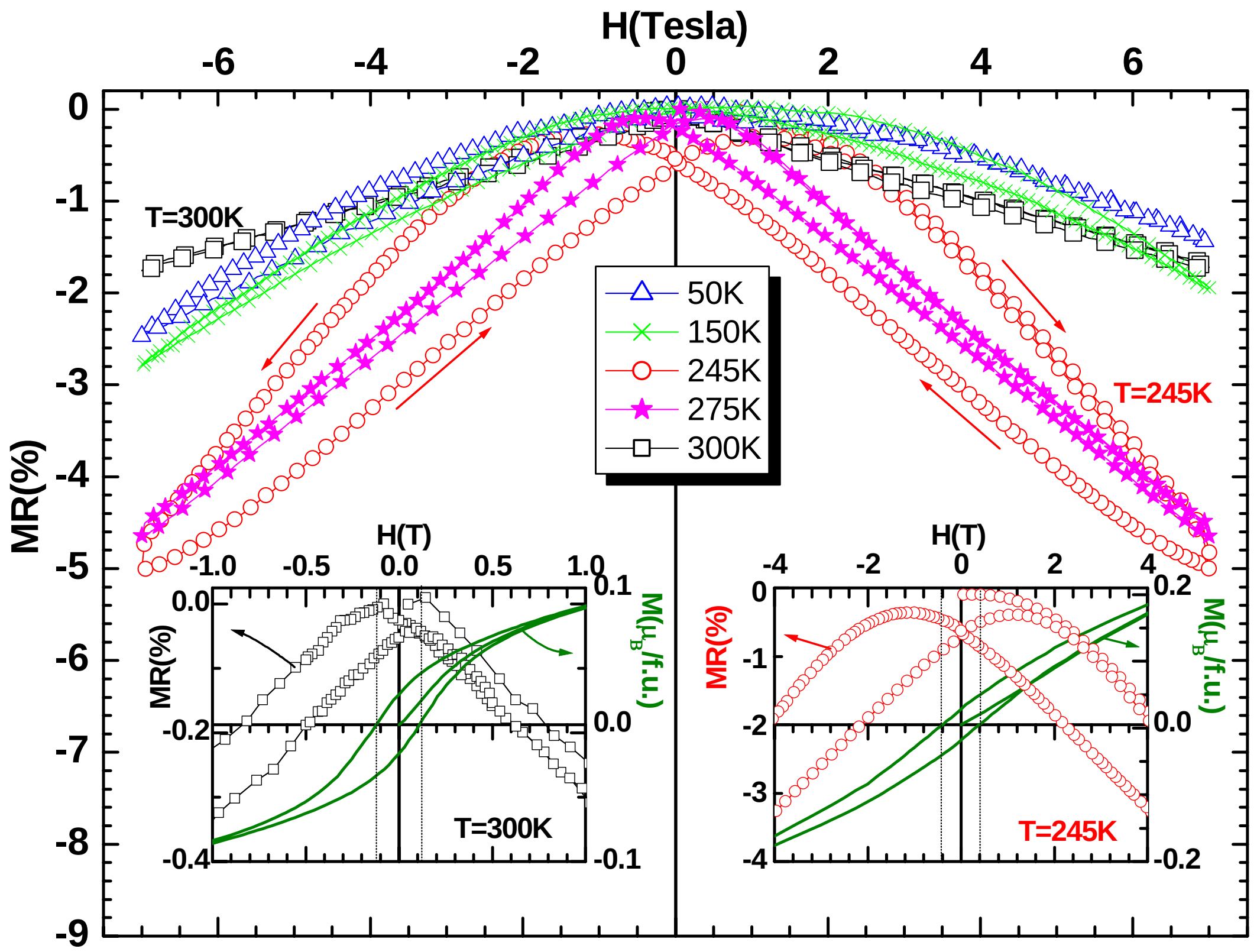

**Figure 3.** Magnetic field dependent isotherm magnetoresistance, MR, effect for ordered $LaBaCo_2O_{5.5}$ at five different temperatures (H= ±7 Tesla). Inset figures show the isotherm magnetization, M(H), and MR plot at 245 and 300 K for comparison, the dotted vertical lines represent the coercive field values.

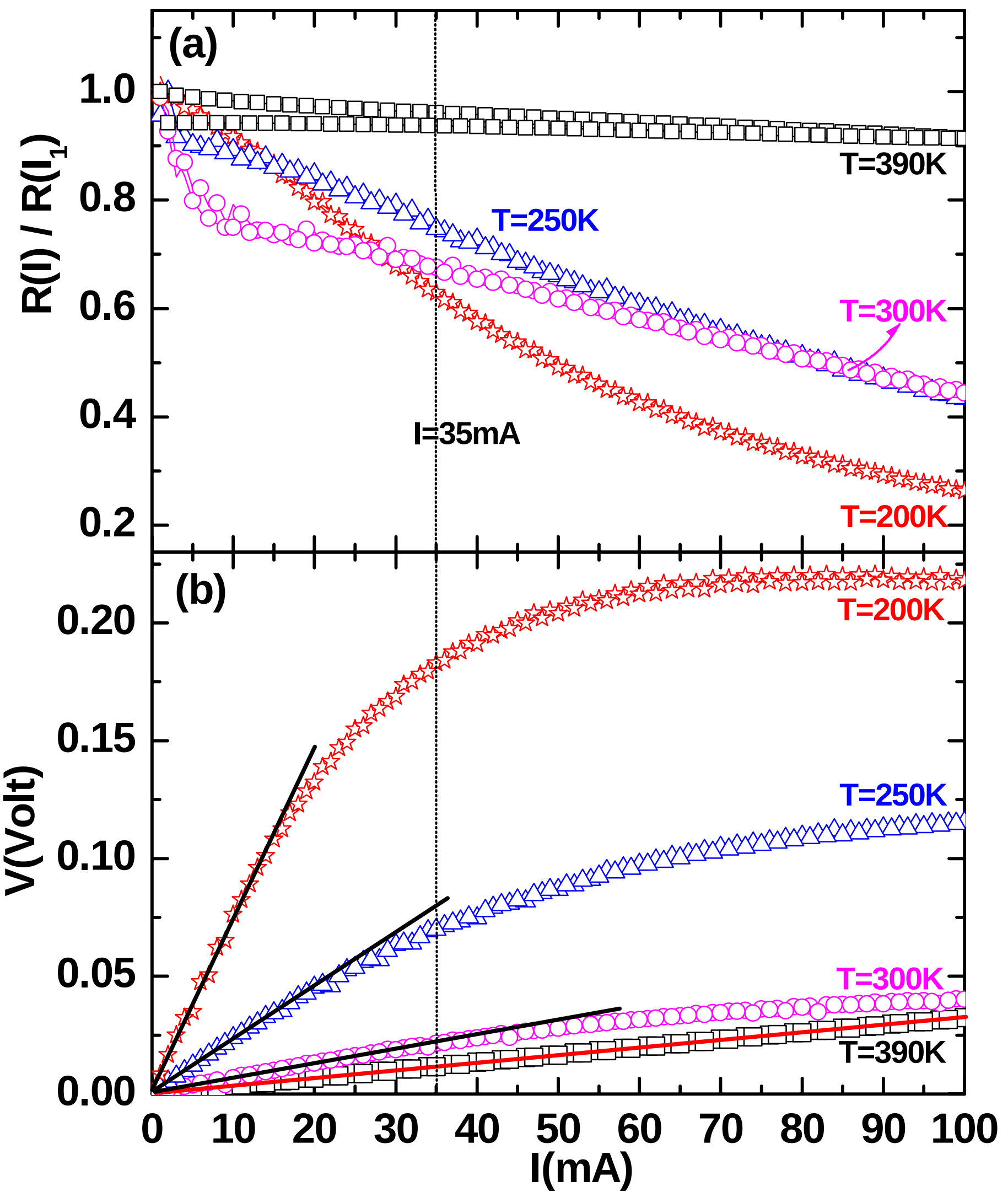

**Figure 4.** (a) The normalized resistance, $R(I)$ / $R(I_1)$; where $R(I)$ is the electrical resistance at 100 mA and $R(I_1)$ for 1mA current, as a function of current bias, and (b) Current (I)-voltage (V) characteristics: open symbol experimental data and solid lines are the linear fit, at four different temperatures for ordered $LaBaCo_2O_{5.5}$.

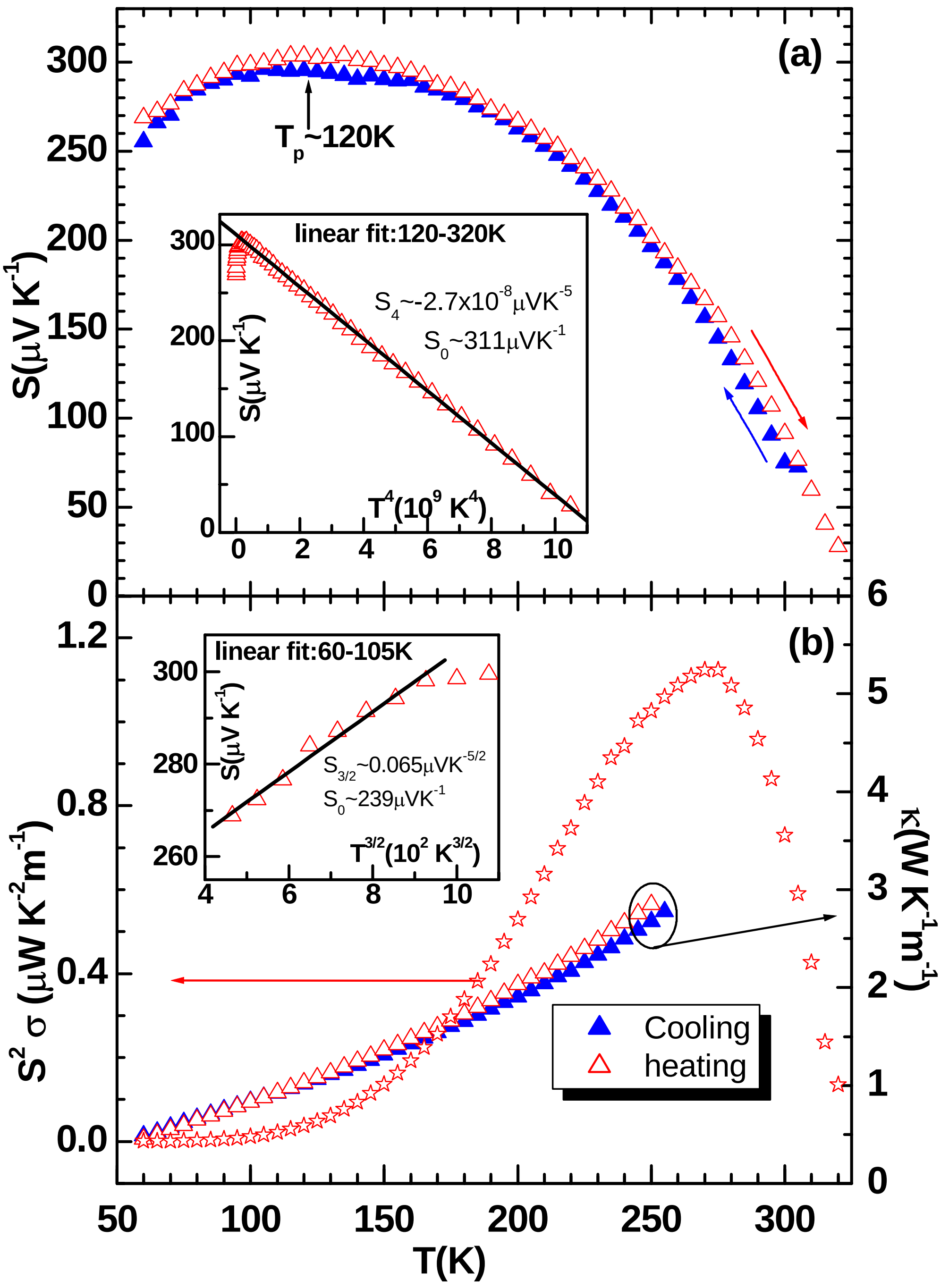

**Figure 5.** Temperature dependent measurements for ordered $LaBaCo_2O_{5.5}$: (a) thermoelectric power, S(T), during cooling (solid triangle) and heating (open triangle) cycles and inset shows the $S\text{-}T^4$ plot as discussed in the text, and (b) thermal conductivity, κ(T), and power factor, $S^2\sigma$(T), plots (inset shows the $S\text{-}T^{3/2}$ plot).